# Advanced Join Patterns for the Actor Model based on CEP Techniques

Humberto Rodriguez Avila[a], Joeri De Koster[a], and Wolfgang De Meuter[a]

a Software Languages Lab, Vrije Universiteit Brussel, Belgium

**Abstract**

**Context** Actor-based programming languages offer many essential features for developing modern distributed reactive systems. These systems exploit the actor model's isolation property to fulfill their performance and scalability demands. Unfortunately, the reliance of the model on isolation as its most fundamental property requires programmers to express complex interaction patterns between their actors to be expressed manually in terms of complex combinations of messages sent between the isolated actors.

**Inquiry** In the last three decades, several language design proposals have been introduced to reduce the complexity that emerges from describing said interaction and coordination of actors. We argue that none of these proposals is satisfactory in order to express the many complex interaction patterns between actors found in modern reactive distributed systems.

**Approach** We describe seven smart home automation scenarios (in which an actor represents every smart home appliance) to motivate the support by actor languages for *five* fundamentally different types of message synchronization patterns, which are lacking in modern distributed actor-based languages. Fortunately, these five types of synchronisation patterns have been studied extensively by the Complex Event Processing (CEP) community. Our paper describes how such CEP patterns are elegantly added to an actor-based programming language.

**Knowledge** Based on our findings, we propose an extension of the *single-message matching* paradigm of contemporary actor-based languages in order to support a *multiple-message matching* way of thinking in the same way as proposed by CEP languages. Our proposal thus enriches the actor-model by ways of declaratively describing complex message combinations to which an actor can respond.

**Grounding** We base the problem-statement of the paper on an online poll in the home automation community that has motivated the real need for the CEP-based synchronisation operators between actors proposed in the paper. Furthermore, we implemented a DSL — called Sparrow — that supports said operators and we argue quantitatively (in terms of LOC and in terms of a reduction of the concerns that have to be handled by programmers) that the DSL outperforms existing approaches.

**Importance** This work aims to provide a set of synchronization operators that help actor-based languages to handle the complex interaction required by modern reactive distributed systems. To the best of our knowledge, our proposal is the first one to add advanced CEP synchronization operators to the — relatively simplistic single-message based matching — mechanisms of most actor-based languages.

## 1 Introduction

The actor model [25] is a well-established paradigm for developing distributed reactive systems. The actor model's concurrency properties are fundamental to achieve the kind of performance and scalability required by such systems in the face of having to handle millions of simultaneous connections [10]. In the original actor model, interaction and coordination of actors are achieved entirely by cleverly encoded asynchronous message exchanges between those actors. Nevertheless, whenever an actor is supposed to execute a method in response to a received *set of messages* (rather than just a single message) with certain characteristics, mainstream actor languages (e.g., Erlang, Scala, Elixir) require developers to encode the characteristics of the set manually. This is illustrated in listing 1 (lines 4, 7, 9). In this example, the actor will react only if a particular sequence of messages ($MsgA \rightarrow MsgB \rightarrow MsgC$) is received. To delay the execution of the reaction code (line 11) until the expected messages have been received in the right order, the actor must manually keep track (lines 5, 8, 13) of previous messages, and validate the progress of subsequent message arrivals in a hard-coded fashion (line 10).

Briefly, the single-message match mechanism of traditional actors complicates the construction of said distributed reactive systems. Developers are forced to manually weave and braid two orthogonal concerns of their actor's interactions and coordinations: when to react (i.e., precisely describe the set of messages that is supposed to give rise to a certain behaviour) and how to react (i.e., the code that describes the actual method to be fired upon reception of said set).

**Listing 1** Example of how to detect a sequence of messages in Elixir

```
1  def loop({ts_a, ts_b}) do
2    state =
3      receive do
4        {:msg_a, timestamp} ->
5          {timestamp, ts_b}
6
7        {:msg_b, timestamp} ->
8          {ts_a, timestamp}
9        {:msg_c, timestamp} ->
10         if ts_b > ts_a do
11           # reaction code
12         end
13         {0,0} # reset state
14     end # receive-end
15   loop(state)
16 end
```

For more than three decades, researchers have been developing new programming language features to improve the expressiveness of the interaction and coordination features of actors. In section 5 we will present an extensive overview of the state-of-the-art in this domain. One notable technique is based on *join patterns*. This technique was popularised by Haller and Van Cutsem [23], where an extension of Scala is presented featuring join patterns. Join patterns where invented by Benton, Cardelli, and Fournet [9] as part of the join calculus. They were added to the thread-based concurrency model of C# leading to a language called Polyphonic C#. The Scala joins patterns of [23] can be seen as an attempt to transpose the mainly synchronous incarnation of join patterns in Polyphonic C# to the asynchronous world of Scala.

Join patterns allow many interaction and coordination patterns to be expressed very elegantly. For example, line 2 in listing 2 exemplifies a join pattern that expresses the coordination between the two methods (`Get` and `Put`) of a `Buffer` class. In this example, the calling thread of the `Get` method will be blocked until the asynchronous `Put` method is invoked. The ampersand (`&`) symbol expresses declaratively that *both* threads need to rendezvous before the method's body is executed.

**Listing 2** Expressiveness of a join pattern in Polyphonic C#

```
1 public class Buffer {
2     public string Get() & public async Put(string s) {
3         return s;
4     }
5 }
```

In this paper, we use seven smart home automations (in which every home appliance is represented by an actor) to motivate the necessity to easily encode very complex interactions and coordinations between actors. Based on an extensive poll in the smart home community on the internet, we show that these are seven real-world concerns that are on the radar of actual developers of such systems. From these seven automations, we identify *five different types of message synchronisation operators*. Then, we show that the current state-of-the-art in actor coordination and interaction technologies in general, and join patterns in particular falls short in supporting these operators.

Based on this problem statement, we present Sparrow, a dialect of Elixir[1] that features actors whose complex interaction and coordination patterns can be described in a highly declarative fashion. Sparrow's interaction patterns (which support our identified synchronisation operators) have been harvested from our extensive literature study of the state-of-the-art of Complex Event Processing (CEP). Hence, Sparrow can be seen as an actor language whose actors have been enriched with CEP and join pattern ideas. We use a smart home scenario as an application domain to help steer our research. However, we conjecture Sparrow as a general-purpose actor coordination framework that can be used to express synchronisation patterns for other event-driven domains (e.g., SmartCities[26], SmartBuildings[43]) as well.

The main contributions of this paper are:

- We showcase current challenges for modern actor languages to coordinate large groups of heterogeneous actors using seven smart home scenarios (see section 2). These use cases are not just synthetic scenarios but have been validated in the smart home community.

[1] Elixir can be regarded as a modern Erlang (e.g., with macros) that runs atop BEAM; i.e., the Erlang virtual machine.

- We present a novel set of advanced actor synchronisation abstractions by formulating them as a domain-specific Elixir extension, called Sparrow (see section 3).
- We validate our proposal by implementing the seven smart home scenarios to demonstrate the expressiveness of the resulting code (see section 4).
- We show the novelty of our approach by contrasting it with the state-of-the-art and related work on actor interaction and coordination techniques (see section 5).

## 2 Motivating Scenarios: Coordination of Smart Home Devices

One application domain where complex interactions between different participants of a system occur is the smart home automation domain. In our scenario, each smart device is digitally represented by an actor (also called *digital twin* [2]). These digital twins can be run on the device itself or any other machine, for instance, on the edge (e.g., raspberry pi) or a cloud-based platform (e.g., SmartThings). The complex interactions between the smart home devices then need to be encoded through message passing between their respective digital twins. For complex scenarios, this will require advanced synchronisation patterns to express these different interactions. In this section, we present seven home automation scenarios that exemplify five fundamental synchronisation operators required to coordinate a group of actors. These are:

1. *Turn on the lights in a room if someone enters, and the ambient light is less than 40 lux.*
2. *Turn off the lights in a room after two minutes without detecting any movement.*
3. *Send a notification when a window has been open for over an hour.*
4. *Send a notification if someone presses the doorbell, but only if no notification was already sent in the past 30 seconds.*[2]
5. *Detect home arrival or leaving based on a particular sequence of messages, and activate the corresponding scene.*[3]
6. *Send a notification if the combined electricity consumption of the past three weeks is greater than 200 kWh.*
7. *Send a notification if the boiler fires three Floor Heating Failures and one Internal Failure within the past hour, but only if no notification was sent in the past hour.*

The above examples were inspired by real automation rules shared on community forums of smart home platforms (e.g., openHAB[4] and Hass[5]). While conducting this study, we only considered complex scenarios that cannot be expressed using a simple mobile app (e.g., Home App from Apple). Each of these complex scenarios was then classified and consolidated into the seven different scenarios described above. These consolidated scenarios highlight different complex coordination patterns between a group of smart devices, which can be abstracted into five primary synchronization operators described at the end of this section. To validate that each of our scenarios is a representative example of the different synchronisation patterns found in the wild, we published an online poll. Each respondent was required to reply with a yes-or-no answer whether or not they had to implement synchronisation patterns similar to each of our seven scenarios.

[2] The postman always rings twice.
[3] A scene is set of actions to be taken in a certain state of the home automation system.
[4] openHAB Forum (https://community.openhab.org, last accessed 2020-10-01).
[5] Hass Forum (https://community.home-assistant.io, last accessed 2020-10-01).

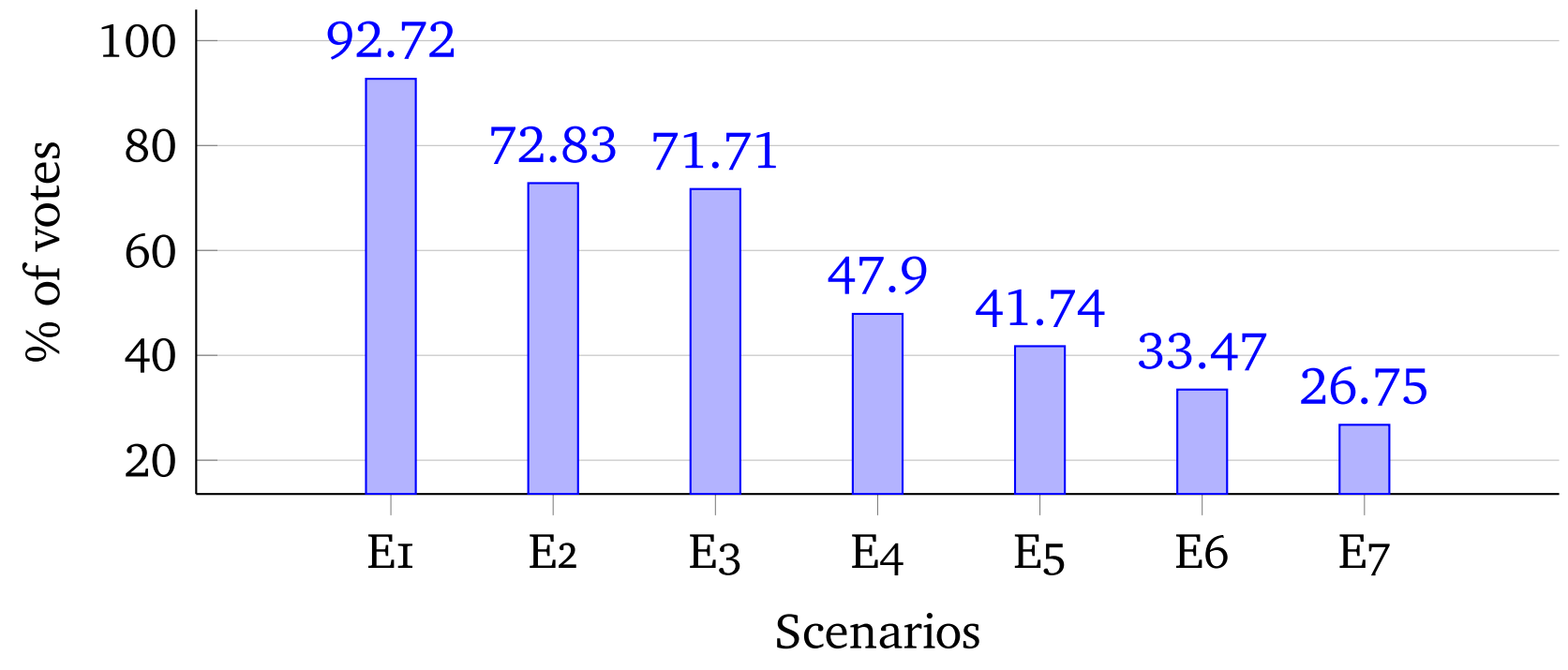

**Figure 1** Results of the online poll. Total of voters: 714. Voting time: 1 month

Our poll was published in four smart home community forums (openHAB [7], Hass [5], SmartThings [8], and Hubitat [6]) for one month and, during that time, collected 714 votes. Figure 1 shows the results of this poll. Due to our poll's non-probabilistic nature, we cannot derive any statistical certainties based on its results. However, it does show that our scenarios are good examples of concerns that present themselves within that community. For example, scenario 1 exemplifies the need for synchronizing multiple independent messages. In this scenario, there are two sensors (motion and light) whose messages need to be synchronized. This concern was identified and raised as a known issue by 662 (92.72 %) of the voters. Scenarios 2, 3, 4, and 5 rely on fairly advanced time-based synchronisation of messages. Furthermore, scenario 5 requires to enforce a particular order in which its constituent messages must match. Our poll shows that timing-constraints and message-order are also two well-recognized concerns within the home automation community. Even the complex scenarios (6 and 7) that require aggregation of multiple messages of the same type over a certain time window were still encountered by more than a quarter of the respondents.[6]

The results of our poll do not validate our list of scenarios as exhaustive. However, it does give us the certainty that the listed scenarios represent common synchronisation patterns found in the wild. Despite these results, we found that both current thread-based smart home platforms and actor languages lack synchronisation abstractions to coordinate heterogeneous smart home devices. We base this statement on our experience after implementing all our examples in two popular smart home platforms

[6] Possibly because scenarios (6, 7) target expensive devices which currently offer limited integration with third-party platforms.

(openHAB and Hass) and a modern actor language (Elixir). These selected programming environments offer at least the same synchronisation abstractions as their state-of-the-art alternatives (e.g., SmartThings, Akka/Scala). The full implementation of each of the seven scenarios using each of the three programming environments can be found online [4]. Each of our implementations using both openHab and Hass was presented and improved based on feedback from their respective communities.

Listing 3 shows the degree of complexity that users of current smart home platforms have to handle by themselves to implement scenario 5. Our openHAB solution considers two motion sensors, one in the entrance hall ($\alpha$), and the second one outside the front door ($\delta$). Furthermore, we use a contact sensor ($\beta$) to detect when the front door was opened. The occupied-home scene is enabled by the following sequence of messages $\delta \rightarrow \beta \rightarrow \alpha$, and the empty-home scene by $\alpha \rightarrow \beta \rightarrow \delta$. Both scenes will be activated if the three messages occur in a time window of 60 seconds. This implementation also assumes that only one person lives in the smart home. Although our solution was improved thanks to the feedback received from the openHAB community, it exposes a lack of programming abstractions to synchronize messages from multiple devices. First, we are responsible for keeping track of each sensor's last update ($\beta$ line 9, $\alpha$ line 15, $\delta$ line 28). Second, we manually discard messages older than the 60 seconds time window (lines 17, 30). Finally, we have to verify that the messages were received in the right order (lines 20, 33) before executing the automation's reaction. In summary, our openHAB solution and its Elixir alternative required more than 20 lines of essential code to express the coordination of only three devices.

Further study of each of the scenarios allowed us to disambiguate five categories of operators that are required to express each of the synchronisation patterns.

**Filter operators** are required to enable the filtering of messages based on their attribute values or timing constraints. For example, in scenario 1, we are required to filter out messages from the ambient light sensor for which the value does not reach 40 lux. Scenario 4 and 6 show the need for filtering messages based on the absence or presence of messages within a certain time window respectively.

**Selection operators** are required to select messages from the set of matching messages after filtering. Traditional actor languages only allow for the consumption of the oldest (first in) message. However, we will require more flexible message selection policies. For example, in scenario 7, we are only interested in the latest *Internal Failure* message, and the third *Floor Heating Failure* message. In general, expressing these types of constraints will require selection operators that allow us to select any message from the set of matching messages after filtering.

**Correlation operators** are required to match a set of different types of messages and to unify their attributes. For example, in scenario 1, we are required to match a motion sensor message and an ambient light sensor message but only match if both are present. Additionally, in scenario 5, we are also required to specify the order in which those messages arrived in order to trigger the right scene. Depending on the order in which sensors detected movement, the person either left the house or arrived home.

**Listing 3** Jython-script implementation for scenario 5 in openHAB

```
c_door = ZDT.now().minusHours(24)
m_hall = ZDT.now().minusHours(24)
m_door = ZDT.now().minusHours(24)

@rule("(Py) Front Door Opened")
@when("Item Front_Door_Contact changed to OPEN")
def front_door_opened(e):
    global c_door
    c_door = ZDT.now()

@rule("(Py) Motion Detected - Entrance Hall")
@when("Item Entrance_Hall_Motion changed to ON")
def entrance_hall_motion(e):
    global m_hall, m_door, c_door
    m_hall = ZDT.now()

    if m_door.isBefore(m_hall.minusSeconds(60)):
        return

    if m_hall.isAfter(c_door) and c_door.isAfter(m_door):
        # code logic for arriving home


@rule("(Py) Motion Detected - Front Door")
@when("Item Front_Door_Motion changed to ON")
def front_door_motion(e):
    global m_hall, m_door, c_door
    m_door = ZDT.now()

    if m_hall.isBefore(m_door.minusSeconds(60)):
        return

    if m_door.isAfter(c_door) and c_door.isAfter(m_hall):
        # code logic for leaving Home
```

**Accumulation operators** are required to aggregate a set of matching messages after filtering. For example, in scenario 6, we are required to aggregate electricity consumption messages of the last three weeks into their sum.

**Transformation operators** are required to transform (e.g., map) a list of accumulated messages. These messages can be subject to new predicate conditions. For example, in scenario 6, we need to check if the total electricity consumption is greater than 200 kWh before sending a notification. Only if the predicate condition is true all the accumulated messages are consumed.

In this section, we have identified the above five types of operators that are required to express complex synchronisation patterns. Unfortunately, as we show in more detail in section 5, state-of-the-art actor languages and frameworks lack the necessary support for these types of operators. Contrary, the above operators are frequently supported by CEP languages/frameworks (e.g., TESLA [11], ETALIS [3], PARTE [37], FlinkCEP [47], and Esper [27]). In the next section, we explore the integration of the well understood CEP operators into actors to overcome these limitations.

## 3 Advanced Join Patterns

In this section, we introduce the main contribution of this paper, to wit a novel join pattern language called Sparrow. Like previous join pattern languages [23, 32, 33], Sparrow extends the traditional single-message match interface of actors [25] to support multiple-message match. However, in order to support the five synchronisation operators described in the previous section, Sparrow's join pattern language design is heavily inspired by CEP languages and frameworks. Sparrow leverages the macro facilities of the Elixir programming language for its implementation. As such, Sparrow is implemented as a domain-specific language (DSL) that benefits from the integration of the extensive set of Elixir/Erlang libraries in that ecosystem. The conceptual model underlying Sparrow's technical implementation was formalised into an executable operational semantics using PLT-Redex [14]. The description of this operational semantics can be found in a companion technical report [40] for this paper.

Listing 4 shows an implementation of scenarios 1 and 2 in Sparrow. A Sparrow module (Actor definition) consists of four parts: importing Sparrow's macros (line 2), pattern definitions (lines 4–-14), reaction definitions (lines 16, 17), reaction bindings (lines 19, 20).

The Sparrow programming language can be subdivided into two smaller languages. On the one hand, there is the *Sparrow's pattern language* (see section 3.1); this is a declarative subset of Sparrow (inspired by CEP languages) that enables the declaration of complex synchronisation patterns (see listing 4, lines 4–-14). On the other hand, the *Sparrow's reaction language* (see section 3.2) is a superset of the Elixir programming language in which the reaction to a matching pattern can be specified (see listing 4, lines 16, 17). The `react_to with:` special form ties both languages together. It determines which reactions will be fired after a complex synchronisation

■ **Listing 4** Main parts of a Sparrow program

```
defmodule SparrowByExample do
    use Sparrow.Actor

    pattern motion as {:motion, id, status, room}
    pattern light as {:light, id, status, room}

    pattern on_motion as motion{status= :on}
                         and light{status= :on}
                         and {:amb_light, id, value, room} when value > 40,
                         options: [last: true]

    pattern no_motion as not motion{status= :on}[window: {2, :mins}]
                         and light{status= :on},
                         options: [last: true]

    reaction turn_on_light(l, i, t), do: # send on command
    reaction turn_off_light(l, i, t), do: # send off command

    react_to on_motion, with: turn_on_light
    react_to no_motion, with: turn_off_light

end
```

```
<pattern-definition>  :=  pattern <identifier> as <pattern>
<pattern>             :=  <elem-pattern> [(and | or ) <elem-pattern>]* [<guard>] [, [options: <option>+]]
<elem-pattern>        :=  [not] <selector> [[<operator>+]] {|> <transformer>}*
<selector>            :=  {<symbol>, <attribute>*} | <identifier>[{{<inline-guard> | <alias-op>}+}]
<attribute>           :=  <value> | <symbol> | <logic-var>
<guard>               :=  when <expression>
<inline-guard>        :=  <identifier> = <expression>
<alias-op>            :=  <identifier> ~> <identifier>
<symbol>              :=  :<identifier>
<logic-var>           :=  [(@ | ! )]<identifier>
<operator>            :=  window: <time> | debounce: <time> | every: <number> | count: <number>
<transformer>         :=  fold(<expression>, <expression>) | bind(<identifier>)
<option>              :=  seq: <boolean> | interval: <time> | last: <boolean>
<time>                :=  {<number>, (:secs | :mins | :hours | :days | :weeks)}
```

■ **Figure 2** EBNF-styled grammar for the Sparrow's pattern language pattern has been detected. The next subsections describe both parts of the Sparrow language in detail and their underlying implementation (see section 3.3). Figure 2 shows an EBNF-styled grammar for the Sparrow's pattern language. The definition for `<expression>`, `<identifier>` `<value>`, `<number>` and `<boolean>` have been omitted and corresponds to ordinary Elixir expressions, identifiers, primitive values, numbers and booleans respectively.

### 3.1 Sparrow Pattern Language

Similarly to the Erlang/Elixir family of actor languages, in Sparrow, incoming messages are *pattern matched* against several message patterns. When a match is found, the matching message(s) are *consumed* from the actor's inbox, and the actor starts a process to react to those messages. This reaction is specified in the Sparrow reaction language (see section 3.2). Sparrow has support for three types of patterns: *elementary patterns* enable the matching of single messages, *composite patterns* enable the composition of multiple elementary patterns, and *accumulation patterns* enable the accumulation and aggregation of multiple messages of the same type.

#### 3.1.1 Elementary Patterns

Elementary patterns are the most basic kind of patterns. They always start with a *selector* that designates a single message followed by an optional operator and an optional guard expression. Defining a new pattern in Sparrow can be done using the `pattern` special form. Figure 3 shows the definition of an elementary pattern that matches when an open window is detected at any location (this is part of the implementation for scenario 3). The `pattern` keyword is used to give a name to the pattern. In Sparrow, patterns are second-class citizens that can be reused and composed to define complex patterns. The `as` keyword is followed by the *selector* of the pattern. Similar to Elixir's messages, a selector in Sparrow is represented as a tuple. Its first element determines the type[7] of message to match, and always has a *unique constant* value (`:window` in our example). The other selector elements are called attributes, and they can be primitive values (e.g., String, Number) or *logic variables*. Logic variables represent a dynamic primitive value that is unknown until the matching of the pattern against a message that sits in the actor's inbox.

```
pattern open_window as {:window, id, :open, location}
```

**Figure 3** Example of an elementary pattern: (1) Primitive to declare a pattern; (2) Assign a name for future references; (3) Define the pattern's selector

As with plain Elixir/Erlang, pattern matching on primitive values can be used to filter messages based on their attribute values. For instance, the pattern shown in our example will only match `:window` messages for which the second attribute (*status*) has as value `:open`. The other attributes (*id, location*) are logic variables.

**Operators** are high-level conditions that further delineate the kind of messages that can be matched by the pattern selector. A selector can optionally be followed by one or more operators enclosed in square brackets. Elementary patterns support three types of operators: *negation*, *debouncing*, and *extensional sequencing*.[8] For example, figure 4 shows an implementation for scenario 2 that uses a negation operator. For this scenario, we want to turn off the lights after two minutes without movement. Detecting the absence of motion events can be implemented in Sparrow by using the negation operator (not) in combination with a timing window.

[7] The type of a selector is always an Elixir *atom* (:).

```
pattern turn_off_light as not {:motion, id, :on, location}[window: {2, :mins}]
                          ❶                                ❷
```

**Figure 4** Implementation of scenario 2 using a negated pattern: (1) Negate the selector definition; (2) Set the time window

Negated selectors must always be combined with a time window (e.g., in seconds, minutes, or hours). Every time the selector matches a new message, the time window is reset. Once the time window expires, the pattern is automatically matched.

Figure 5 shows an implementation of scenario 4 using the debouncing[9] operator. For this scenario, we only want to match doorbell messages if no other doorbell message was matched in the past 30 seconds. A debounced operator does exactly that, the first message that matches the preceding selector automatically matches the entire pattern. Any future messages that follow within the debouncing time are automatically discarded.

```
pattern doorbell_alert as {:doorbell, id}[debounce: {30, :secs}]
```

**Figure 5** Implementation of scenario 4 using a debouncing time between messages

Finally, figure 6 presents the use of the extensional sequencing operator in order to implement a partial solution for scenario 7. For our partial solution, we are interested in matching every third heating failure event. This can be done in Sparrow by following the pattern selector with an extensional sequencing operator using the every keyword. In our example, only every third message is matched and consumed. All other messages that match the selector are discarded.

```
pattern heating_failure as {:heating_f, id, code}[every: 3]
```

**Figure 6** Example using the extensional sequencing operator (every) of Sparrow

**Guards** are boolean predicates that are executed after a match of the selector and its operators is found. Like with traditional Elixir guards [48], Sparrow guards are only allowed to contain boolean predicate expressions that can always be executed in constant time (and are side-effect free). Figure 7 shows a guarded pattern that will be activated if it matches an open window message from the bedroom or the kitchen. The full pattern is only matched (and the corresponding reaction fired) when the guard expression evaluates to true. If the guard expression evaluates to false, the pattern is not matched, and the messages are not consumed.

[8] The integration of Sparrow in Elixir required us to design negation as a prefix operator, and debouncing and sequencing as postfix operators.

[9] The term *debouncing* is taken from the domain of electrical circuits where a particular debouncing algorithm is used to avoid multiple triggers (within a period) to produce an undesired control output [19].

```
pattern open_window as {:window, id, :open, location}
                    when location == :bedroom or location == :kitchen
```

■ **Figure 7** Example of a pattern with a guard expression

### 3.1.2 Composite Patterns

Elementary patterns that have been named by means of the pattern keyword can be reused, further specified or composed with other patterns.

**Reusing elementary patterns** can be done by specifying a pattern name instead of a selector when defining a new pattern. An optional set of *inline guards* can follow this name. Inline guards are just syntactic sugar to write compact guard expressions. Pattern reuse is illustrated in figure 8, which shows two semantically equivalent variants (B, C) of a pattern that extends another elementary pattern (A). Both patterns are matched whenever a window of the kitchen is open. Pattern B uses a guard expression to further specify that the open window needs to be detected in the kitchen using the equals operator (==). On the other hand, pattern C employs an inline guard to substitutes the logic variable location for the atom :kitchen using the match operator (=).

```
(A) pattern open_window as {:window, id, :open, location}

(B) pattern kitchen_window_a as open_window when location == :kitchen

(C) pattern kitchen_window_b as open_window{location= :kitchen}
```

■ **Figure 8** Examples of pattern reuse: (A) Definition of an elementary pattern; (B) Further specifying an existing pattern with an additional guard expression; (C) Definition of a reused pattern with an additional inline guard

**Composing patterns** can be done by linking multiple elementary patterns by means of a logic operator. Sparrow supports both conjunctions (and) and disjunctions (or) of patterns.[10] Figure 9 illustrates the use of composing patterns in order to partially implement smart home scenario 5. For this scenario, we are interested in detecting when the user is arriving home by first detecting motion at the front door, followed by receiving an open door event, followed by detecting motion in the entrance hall. In our example, the occupied_home pattern is defined as a conjunction of three elementary patterns. The first and third one reuse the motion_sensor pattern to match only motion events from the *front door* and *entrance hall* respectively, On the other hand, the second one is an anonymous pattern that matches door opening events.

[10] Like with traditional actor languages; disjunction can also be achieved by separating each of the patterns in the disjunction. However, as Sparrow also has conjunction, which is usually not supported by traditional actor languages, we also syntactically support disjunction.

Any composed pattern can be followed by an optional set of operators. By default, a composed pattern does not enforce any particular order in which its constituents' patterns must match. However, that behaviour can be changed using the *intensional sequencing* operator (seq). In our example of figure 9, we set sequencing to true in order to specify that the occupied_home pattern can only be matched if the matched messages arrive in the same order as they are specified in the pattern definition.

```
pattern motion_sensor as {:motion, id, :on, location}

pattern occupied_home as motion_sensor{location= :front_door}
                      and {:contact, id, :open, :front_door}
                      and motion_sensor{location= :entrance_hall},
                      options: [ seq: true ]
```

**Figure 9** Example of use of sequencing operator (seq)

**Renaming logic variables** Sparrow has support for unification of logical variables crossing the constituents of a composed pattern. This is often desirable as it allows for the unification of various attributes across different pattern selectors. However, Sparrow currently unhygienically expands all named patterns into the composed pattern (similar to unhygienic macros). This can potentially lead to unexpected unification of logical variables when composing *named* patterns. For example, our first implementation of scenario 5 in figure 9 contains a bug as it incorrectly does not match the sequence of messages shown in figure 10.A. Figure 10.B shows the expanded form of the composite pattern defined in figure 9. Each of the selectors of the elementary patterns contains the same logical variable id. However, in this case, it is undesirable to unify these three logical variables as each sensor can have a different id. To circumvent this issue, Sparrow allows developers to manually change the logical variable for an attribute using the aliasing operator (~>). The aliasing operator renames the logical variable on its left-hand side to the logical variable on its right-hand side in the elementary pattern. This operator is not an optimal solution. In future versions of Sparrow, the default behaviour of unification will be changed to facilitate the maintenance of large pattern sets. However, as patterns are not first-class entities, developers can identify shared logic variables by looking at the pattern definition. Figure 10.C presents an improved version of the occupied_home pattern.

**Timing constraints on composite patterns** Similar to windowing for elementary patterns, composite patterns also support timing-constraints. Developers can specify a *time-interval* in which the composed set of patterns should be matched. Figure 11 presents a new version of the occupied_home pattern where the 60 seconds time constraint is added. Similar to intentional sequencing, this time constraint (interval) is also defined in the options of a composite pattern.

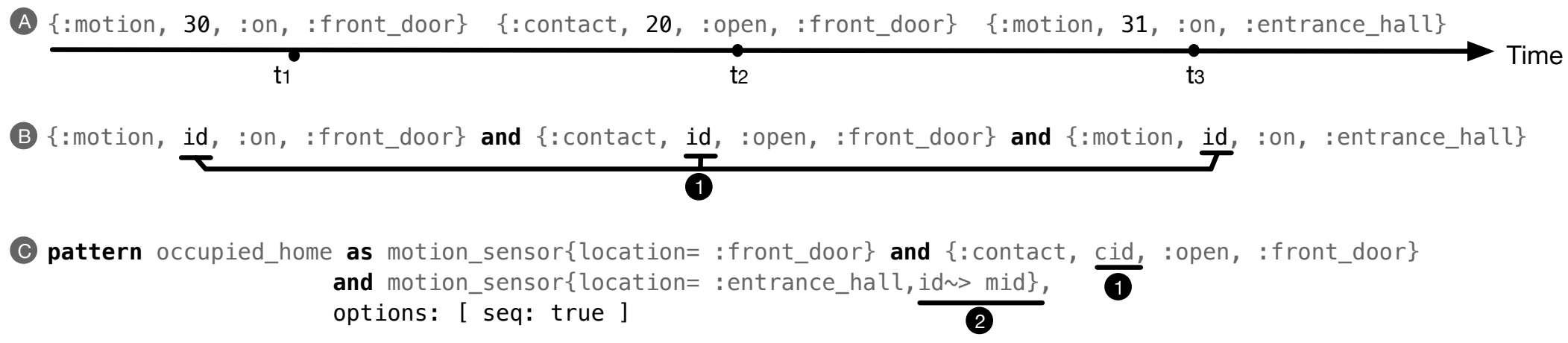

**Figure 10** Example of a composite pattern with shared logic variables: (A) Sample of messages; (B) Expansion of the occupied_home pattern's composite selector; (B.1) Highlight shared logic variables; (C) Fixed declaration of the occupied_home pattern; (C.1) Change attribute identifier; (C.2) Rename attribute identifier using the alias operator

```
pattern occupied_home as motion_sensor{location= :front_door}
                      and {:contact, cid, :open, :front_door}
                      and motion_sensor{location= :entrance_hall,id~> mid},
                      options: [ seq: true, interval: {60, :secs} ]
```

**Figure 11** Example of a composite pattern with a time interval constraint

**Composite patterns selection strategy** Like most actor languages, Sparrow messages are matched in FIFO order. However, Sparrow enables one to deviate from that default selection strategy. Figure 6 already showed an example of this by only selecting every third message. This particular message selection strategy is useful to synchronize always on the latest messages that may be relevant to a pattern. For example, the implementation of the pattern occupied_home from figure 11 must always check all potential messages received from the three sensors in the last 60 seconds. However, as observed in figure 12.A at $t_4$, the pattern should only check the latest message ($t_3$) from the entrance hall's motion sensor and discard the old ones ($t_1, t_2$).

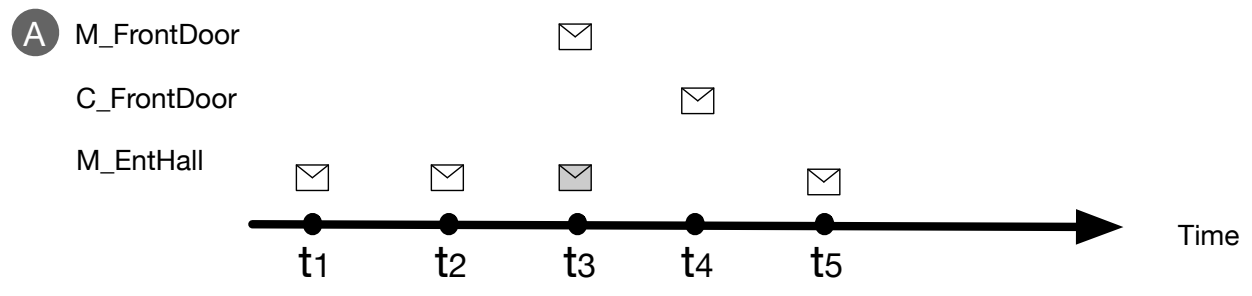

```
B pattern occupied_home as motion_sensor{location= :front_door}
                        and {:contact, id, :open, :front_door}
                        and motion_sensor{location= :entrance_hall, id ~> mid},
                        options: [ seq: true, interval: {60, :secs},last: true ]
```

**Figure 12** A solution to the occupied-home scene of scenario 5: (A) Messages received by the actor; (B) Composite pattern that enforces a selection strategy (*last-in*)

### 3.1.3 Accumulation Patterns

Accumulation patterns extend elementary patterns with quantified and unquantified accumulation of sets of the same type of message. Once the set of messages is accumulated, further transformation and filtering can be applied. Accumulation patterns can be constructed from elementary patterns by means of several optional accumulation operators. Like other operators (see section 3.1.1), they can be specified between square brackets following the selector.

**Quantified accumulation patterns** are used to accumulate a fixed number of matching messages. Figure 13.A shows a pattern that uses the count operator to match three heating failure messages. However, similar to reused patterns, the pattern expansion step of Sparrow's macros will unhygienically expand these patterns. In our example, this means that the pattern heating_failure accumulates three heating failure messages from the same boiler and with the same failure code as the logical variables id and code will be unified (see figure 13.B). Although this default behaviour may be useful in some circumstances, it is not always desirable.

Sparrow introduces the operators: *must be distinct* ! and *may be distinct* @, to specify the expected matching behaviour of logic variables when used in an accumulation pattern. The former guarantees that the constrained attribute must have a *distinct* value in *all* the messages accumulated. In contrast, the latter allows the constrained attribute to have *any* value. For example, the definition of heating_failure pattern shown in figure 13.C will match three messages from the same boiler regardless of the *code* attribute's value.

```
(A) pattern heating_failure as {:heating_f, id, code}[count: 3]

(B) {:heating_f, id, code} and {:heating_f, id, code} and {:heating_f, id, code}

(C) pattern heating_failure as {:heating_f, id, @code}[count: 3]
```

**Figure 13** Example of a quantified accumulation pattern that matches three heating failure messages: (A) Use of the operator count to accumulate tree messages; (B) Highlight shared logic variables in expanded pattern; (C) Use of the *may be distinct* operator @ to accumulate messages regardless of the *code* attribute's value

**Unquantified accumulation patterns** are used to accumulate any number of messages within a certain time window using the window operator. Figure 14.A shows a pattern that uses the window operator to accumulates all heating failure messages from the last 60 minutes. Messages older than the time constraint are automatically removed by Sparrow's run-time. As a second example of unquantified accumulation patterns, figure 14.B shows a hybrid accumulation pattern that use both count and time-based accumulation operators. In this last case, the first operator that reaches its condition will win and result in a match.

**Transformation operators** allow developers to transform and filter a group of received messages that have been accumulated. For example, figure 15.B presents a pattern that implements the requirements for scenario 6. Notice that the electricity_alert pattern uses the *may be distinct* operator (B.1) to match all daily consumption values. Once the accumulation operator is satisfied (B.2), the list of messages is transformed (B.3). Later, the pattern uses the special form bind (B.4) to save total electricity consumption

```
A pattern heating_failure as {:heating_f, id, @code}[window: {60, :mins}]

B pattern heating_failure as {:heating_f, id, @code}[count: 3, window: {60, :mins}]
```

**Figure 14** Examples of unquantified accumulation patterns: (A) Use of the window operator to accumulate all messages in the last 60 minutes; (B) Example mixing both accumulation operators

in a local variable (*total*). Finally, the guard expression (B.5) is evaluated to determine if the messages are consumed or not by the pattern. In this example, messages older than three weeks are automatically discarded by the Sparrow runtime.

```
A pattern daily_electricity as {:consumption, meter_id, value}
                                                    1         2
B pattern electricity_alert as daily_electricity{@value}[window: {3, :weeks}]
                               |> fold(0, fn({_,_,v}, acc)-> acc + v end)     3
                               |> bind(total)                                 4
                               when total > 200                               5
```

**Figure 15** Sparrow solution for scenario 6: (A) Elementary pattern definition; (B) Accumulation pattern example with a transformer operator and guard

### 3.2 Sparrow Reaction Language

So far, we have only focused on the Sparrow's pattern language. However, once a pattern is matched, the matched messages are consumed, i.e., removed from the actor's inbox, and the actor starts reacting to the matched pattern. In Sparrow, this reaction logic is syntactically decoupled from the pattern definition (similar to [51]). A reaction can be dynamically bound to one or more patterns, and a pattern can have multiple reactions. This behaviour was intentionally designed to facilitate the reuse of both patterns and reactions. Furthermore, it avoids developers to add an extra layer of indirection (middleman) to determine which reactions to execute. The middleman approach circumvents the duplicity definition of both patterns and reactions. However, it also introduces a performance penalty since it is always notified whether a pattern has registered reactions or not. In summary, the decoupling of patterns and reactions allows the programmer to dynamically change the behaviour of a Sparrow actor, which is reminiscent of a *become* statement in the original actor model.

Figure 16 showcases a smart home scenario where the binding of reactions to a pattern is based on the current season of the year (e.g., summer, winter). In this way, developers can avoid duplicated patterns with different reactions to match the season requirements. This figure omits the code related to the dynamic scheduling of the reactions. Figure 16.A shows the definition of two reactions (turn_off_heating, turn_off_cooling) using the reaction special form. Both reactions can be bound to the pattern window_open (see figure 16.B) based on the current season. The definition of a reaction is similar to the definition of a named function that always has the same

```
A reaction turn_off_heating (mgs, int_res, state), do: # reaction code
                              1      2       3             4
  reaction turn_off_cooling (mgs, int_res, state), do: # reaction code

B pattern open_window as {:window, id, :open, location}

C react_to open_window, with: turn_off_heating
               1                    2

D remove turn_off_cooling, from: open_window
                1                    2

E remove_reactions open_window
```

**Figure 16** Overview of reaction primitives: (A) Define two reactions; (B) Add a reaction to a pattern; (C) Remove a reaction from a pattern; (D) Remove all reaction of a pattern

three parameters. The first one (A.1) is a list of messages matched by the pattern. The second parameter (A.2) is a key-value list with all the intermediate results saved with the bind operator during a transformation process (see figure 15.B.4). Finally, the third parameter (A.3) represents the current state of the actor. In this example, the definition of both reactions omits the code related to their body and their dynamic scheduling.

Binding a reaction to a pattern is done using the react_to special form (see figure 16.C) which expects two arguments: the *pattern's name* (C.1), and the reaction's name (C.2). When the pattern open_window (see figure 16.B) is successfully matched, the actor will invoke all its reactions in the same order they were bound to the pattern. A reaction can also be unbound from a pattern. The remove primitive function (see figure 16.D) also expects the same two arguments as the react_to primitive function but in reverse order. Finally, the remove_reactions function (see figure 16.E) removes all the reactions of the pattern received as argument.

### 3.3 Implementation

As mentioned in section 3, Sparrow extends the traditional single-message match interface of actors to support multiple-message match. To achieve that, a Sparrow actor has a virtual-inbox that it is used as a knowledge base of an embedded *pattern engine*. Like the traditional actor's inbox, the virtual-inbox is theoretically unbounded in size. However, unlike traditional actor languages, Sparrow messages have a finite lifetime. Each actor can define a default lifetime for the messages it receives. After a message expires, it is automatically garbage collected by the pattern engine.

The pattern engine builds a directed graph (also known as discrimination network) representing all patterns defined in its actor. Internally, Sparrow's patterns are represented by a special type of node called *pattern node*. These nodes implement the different synchronisations operators motivated in section 2 and supported by Sparrow. Additionally, each pattern node maintains a history (called *buffer*) of previous matched messages. Although in figure 17 patterns are represented by a single type of node,

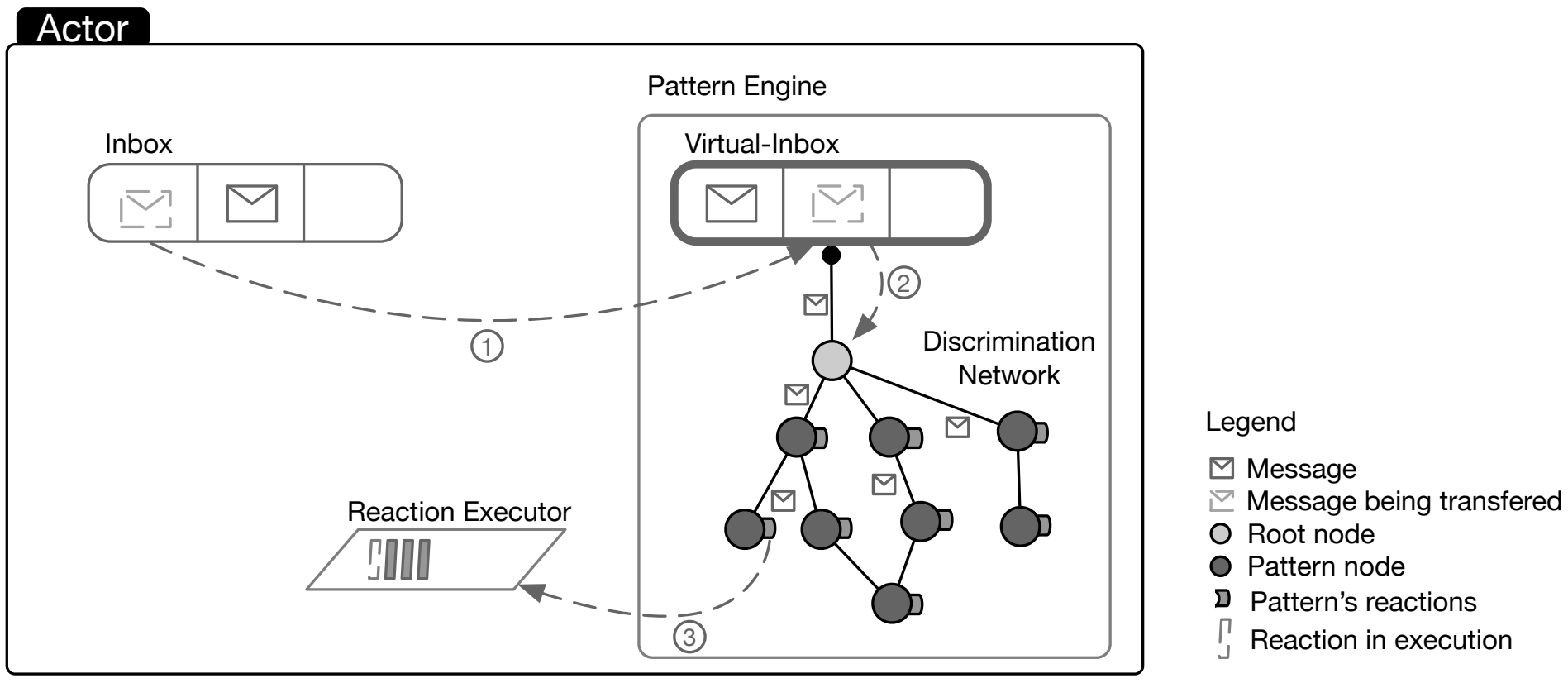

**Figure 17** Overview of the internal representation of a Sparrow actor: (1) Transfer of a received message to the virtual-inbox; (2) Feed the engine's discrimination network with a new message; (3) Queue pattern reaction after a successful match

several subtypes exist, each of them addressing a particular type of pattern defined in section 3.

Figure 17 shows a simplified view of the internal representation of an actor in Sparrow and its message matching process. The matching process starts by transferring each received message from the actor's inbox into the *virtual-inbox*. Later the pattern engine tries to match each new message against a group of patterns for which it is relevant; we call this process a *match-cycle*. The discrimination network's *root node* serves as the entry point of new messages for the matching process. This node will determine potential pattern nodes based on the message's type and will forward it to them. After this step, the message will flow through the discrimination network until a pattern node with a successful match of its conditions is found. In that case, the message is consumed by its reaction(s), which will be one after the other executed sequentially by the actor's *reaction executor*. Otherwise, the message remains in the node's buffer until a successful match is completed or until the message expires. The above process is an essential tool for the engine's incremental matching strategy. This strategy is based on a custom implementation of the RETE algorithm [15]. Furthermore, it implements a *single pattern selection* and *selected message consumption* policies [54]. The former guarantees that a pattern matches at most once per match-cycle. The latter guarantees that a pattern can consume a message only once. However, multiple patterns can consume the same message. The details of this implementation can be found in our GitHub repository.[11]

[11] https://github.com/softwarelanguageslab/sparrow

## 4 Evaluation

In this section, we evaluate Sparrow's expressiveness by comparing its solutions to our seven scenarios against the ones expressed in two popular smart home platforms (openHab, Hass) and one actor language (Elixir). We chose Elixir because it is the host language of our DSL, and it provides synchronisation abstractions similar to other mainstream actor languages (e.g., Erlang, Scala). We followed a quantitative evaluation approach based on an analysis of the lines of code required to express each automation's coordination logic. In other words, our comparison does not consider code not related to the coordination process (e.g., imports, reaction logic of the automations). To obtain a fair comparison between our actor-based solutions and the ones from the smart home platforms, we published these solutions on both community's forums (see openHAB [39] and Hass [38] topics). This allowed us to get feedback, and incrementally arrive at a solution that could be implemented by experts in these communities in our comparison. Figure 18 shows a solution to scenario 5 in openHAB (A), Elixir (B), and Sparrow (C). We decided to analyse this particular solution since it showcases most of the actions that developers have to take care of during the implementation of our scenarios. Due to the similarity between the solutions of openHAB (Jython) and Hass (Python), we omitted the latter. However, the interested reader can inspect this and the rest of our solutions in our GitHub repository [4].

As can be observed in figure 18, for this particular automation, the Sparrow's solution is the most compact of all three. However, we want to compare the *relative* amount of code that we (as developers) have to write for handling all different concerns related to the synchronisation of messages. To do this analysis, we highlighted four concerns that we frequently found in the implementations of our smart home scenarios. The first one (*state management*) indicates the code that is used to save temporal data required by the ongoing coordination process. For example, in both openHAB and Elixir implementations, we had to manually track each sensor's most recent message, spending 42.86 % and 40 % respectively of the coordination code on that action. However, Sparrow shields developers from this kind of work. The second concern (*windowing management*) highlights the code needed to discard messages that do not satisfy the pattern's timing constraints. In contrast, the third one (*sequencing control*) points the code to enforce a particular message order. The coordination code required for these two actions in automation 5 was relatively simple. However, unlike Sparrow (lines 10, 13), the complexity of these synchronisation operations in openHAB (lines 21, 24, 33, 36) and Elixir (lines 8, 9, 16, 17) is directly proportional to the number of devices involved. The last concern, *pattern definition* emphasizes the code used to express the type of messages to be synchronized and their content-based conditions. Although it is the most crucial concern for the coordination process, openHAB and Elixir's patterns only define when to react to single messages. The inability to express the whole coordination process by means of declarative patterns forces developers to shift their focus from *when* to react to *how* to do it. By contrast, in Sparrow (see figure 18.C), a developer focuses on the declaration of patterns, and he lets the run-time to figure out how to match its constituents.

A

```python
from core.rules import rule
from core.triggers import when
from java.time import ZonedDateTime as ZDT

c_door = ZDT.now().minusHours(24)
m_hall = ZDT.now().minusHours(24)
m_door = ZDT.now().minusHours(24)

@rule("(Py) Front Door Opened")
@when("Item Front_Door_Contact changed to OPEN")
def front_door_opened(event):
  global c_door
  c_door = ZDT.now()

@rule("(Py) Motion Detected - Entrance Hall")
@when("Item Entrance_Hall_Motion changed to ON")
def entrance_hall_motion(event):
  global m_hall, m_door, c_door
  m_hall = ZDT.now()

  if m_door.isBefore(m_hall.minusSeconds(60)):
      return

  if m_hall.isAfter(c_door) and c_door.isAfter(m_door):
      # code logic for arriving home

@rule("(Py) Motion Detected - Front Door")
@when("Item Front_Door_Motion changed to ON")
def front_door_motion(event):
  global m_hall, m_door, c_door
  m_door = ZDT.now()

  if m_hall.isBefore(m_door.minusSeconds(60)):
      return

  if m_door.isAfter(c_door) and c_door.isAfter(m_hall):
      # code logic for leaving home
```

B

```elixir
defmodule Automation5 do
  import Timex

  def loop({m_door, m_hall, c_door}) do
   state =
     receive do
       {:motion, _id, :on, :front_door, m_door_dt} ->
         if before?(shift(m_door_dt, seconds: -60), m_hall)  do
           if after?(m_door_dt, c_door) and after?(c_door, m_hall) do
             # code logic for leaving home
           end
         end
         {m_door_dt, m_hall, c_door}

       {:motion, _id, :on, :entrance_hall, m_hall_dt} ->
         if before?(shift(m_hall_dt, seconds: -60), m_door)  do
           if after?(m_hall_dt, c_door) and after?(c_door, m_door) do
             # code logic for arriving home
           end
         end
         {m_door, m_hall_dt, c_door}

       {:contact, _id, :open, :front_door, dt} ->
         {m_door, m_hall, dt}
     end

   loop(state)
  end

end
```

C

```elixir
defmodule Automation5 do
  use Sparrow.Actor

  pattern motion as {:motion, id, :on, location}
  pattern m_front_door as motion{location= :front_door}
  pattern m_entrance_hall as motion{location= :entrance_hall, id~> mid}
  pattern c_front_door as {:contact, cid, :open, :front_door}

  pattern occupied_home as m_front_door and c_front_door and m_entrance_hall,
                          options: [ interval: {60, :secs}, seq: true, last: true ]

  pattern empty_home as m_entrance_hall and c_front_door and m_front_door,
                          options: [ interval: {60, :secs}, seq: true, last: true ]

  reaction activate_home_scene(l, i, t), do: # code logic for arriving home
  reaction activate_leave_scene(l, i, t), do:  # code logic for leaving home

  react_to occupied_home, with: activate_home_scene
  react_to empty_home, with: activate_leave_scene

end
```

State management | Pattern definition | Windowing management | Sequencing control

**Figure 18** Solution for scenario 5 in openHAB (A), Elixir (B), and Sparrow (C)

| | openHAB | Hass | Elixir | Sparrow |
|---|---|---|---|---|
| State management | 29 | 30 | 55 | 0 |
| Sequencing control | 2 | 2 | 2 | 2 |
| Windowing management | 20 | 15 | 12 | 7 |
| Pattern definition | 24 | 13 | 31 | 16 |
| Total lines of code (LoC) | 75 | 60 | 100 | 25 |

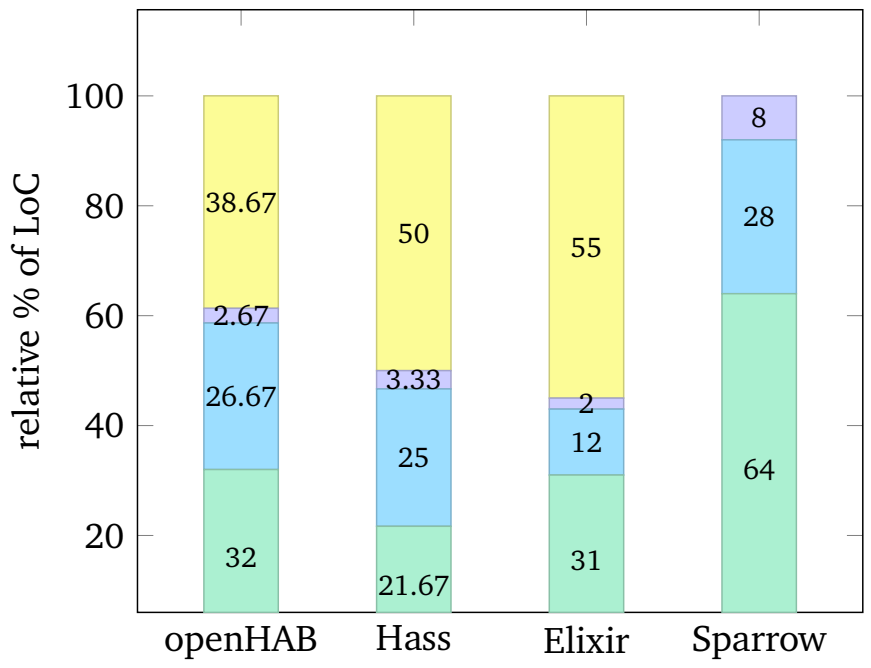

**Figure 19** Summary of analyzing different solutions for the seven scenarios

Figure 19 summarizes our solutions with a breakdown of the total lines of code (LoC) and their respective relative percentage according to the four identified concerns. For each concern and platform/language, we show the sum of all LoC written in our seven solutions. A detailed breakdown for each solution can be found in appendix A. To populate the values shown in this figure, we manually tagged each LoC related to the four concerns. Later, we retrieve their sum values using a helper script (statistics.exs) located in the root directory of each solution group (e.g., openHAB, Sparrow). Using this comparison, we show that Sparrow seems to succeed at reducing the effects of non-functional concerns that arise during the synchronisation of messages in a complex actor system.

## 5 Related Work

Our main goal in this paper was to retrofit CEP operators onto the actor model to improve the expressiveness of encoding the interaction and coordination of actors through join patterns. In this section, we discuss how the related work done in this domain relates to the synchronisation operators described in section 2. We group the below proposals in three main categories: *communication model extension*, *monitor & verification*, and *local synchronisation*. Due to the not mutually exclusive definition of our categories, some proposals may fit in more than one. Table 2 in appendix B summarizes the support for the five types of synchronisation operators by state-of-the-art actor frameworks and languages.

**Communication model extension** Most of the related work in this category add support for a multi-cast message communication [1, 12, 20, 21, 22, 45]. Other related work adds an extra synchronisation layer to the traditional actor's point-to-point communication model [28, 52]. Within this category we can further disambiguate between approaches that synchronize messages on the sender-side and approaches that synchronize messages on the receiver-side. Sparrow falls within the latter category. All of these approaches filter messages based on their values and do not have support for time-based filtering of messages. Moreover, with the exception of interActors [21] and AErlang [12], all of the synchronisation mechanisms in this category are only able to match individual messages. In interActors, the sender of a message can wait for two or more reply messages before continuing. On the other hand, AErlang has support for receiver-side synchronisation by allowing the receiver actor to accumulate and filter a finite number of messages of the same type. Our proposal takes this further by allowing the interface of an actor to synchronize different types of messages. Furthermore, our patterns extend the single message transformation of [28] to a list of messages. Contrary to Sparrow, none of the proposals in this category provide abstractions to detect the absence of messages, or enforce a matching order, or enable matching a disjunction of messages using a single pattern.

**Monitor & verification** Related work in this category [13, 16, 17, 24, 29, 30, 31, 35, 36, 44, 46] uses reflection to observe and limit the interaction of one or a group of actors. The coordination process is mostly done by a special type of meta-actors which enforces a particular protocol for incoming and outgoing messages. In contrast, our proposal is based on a local synchronization of messages in the receiver actor. Like Sparrow, synchronization abstractions of the above proposals can filter messages based on their values and time constraints (e.g., [29, 35, 46]). However, the abstractions provided by [24, 31] only supports type-based constraints. Furthermore, only [35] can detect the absence of messages, but it assumes that actors have their local clocks synchronized, and their invocations are scheduled atomically. Additionally, like traditional actors, the synchronisation abstractions of these proposals always match the oldest messages. Contrary to Sparrow, message correlation operators are used only to enforce a particular communication protocol. The message receiver behaves like a traditional actor matching a single message.

**Local synchronisation** Proposals in this category can be further classified into two main groups. The first one targets synchronisation approaches based on promises/-futures or message-passing continuations [18, 34, 49, 50, 53]. Unlike Sparrow, their abstractions allow the synchronisation and chaining of individual message invocations. The second group extends the actor interface to match a set of messages instead of individual ones. Sparrow, like the other proposals in this category [23, 32, 33] expands the matching capabilities of the traditional actor's receive primitive with *join patterns* [9]. Join pattern languages commonly only support the unification of individual messages, except Activators [18] and Sparrow, where patterns can define disjunctions. However, Activators' patterns cannot mix conjunctions and disjunctions of messages in a single pattern like in Sparrow. Furthermore, unlike Sparrow, joins in none of these languages enforce a particular matching order of their constituents' messages. Languages in this group use pattern-matching techniques to compact filtering expressions. Despite that, only JErlang [33], JCThorn [32], and Sparrow support a non-linear pattern matching mechanism. This mechanism allows join patterns to synchronize the values of shared attributes among multiple messages without guard expressions. The above join languages also force developers to statically bound patterns and reactions during the definition of a pattern. Furthermore, they lack support for timing constraints and accumulation operators to filter messages.

## 6 Conclusions

Complicated message exchange patterns play an important role in the development of modern actor-based applications. We introduce five novel types of synchronisation operators to describe coordination between actors in a modern actor-based language. The operators were inspired by well-established abstractions in the CEP domain. The need for the operators was supported by real-world scenarios coming from the smart home automation community. A poll ran in the said community had endorsed our proposal with fairly large evidence. Even the most exotic operator was on the wish list of no less than 25 percent of the developers. Our proposal has been realised technically in the form of an Elixir dialect called Sparrow. It was validated by implementing the scenarios in Sparrow, in two popular smart home automation platforms (openHab, Hass) and in raw Elixir. By labeling each line of code with the concern it contributes to, we have demonstrated that Sparrow does what it claims to do, namely giving actor programmers the abstractions necessary to declaratively specify complex interaction patterns between actors. Sparrow's join patterns have been implemented using discrimination networks and a variant of the RETE algorithm in order to support performance. At the time of writing, we are benchmarking Sparrow against other actor-based coordination approaches. New versions of our DSL will also mitigate the undesired unification of pattern attributes due to the current unhygienic expansion approach. We are also working on an integration of Sparrow into the Hass smart home platform. Furthermore, we plan to open-source Sparrow to the Elixir/Erlang community, allowing us to harvest more user feedback from both communities.

**Acknowledgements** We are thankful to the anonymous reviewers of Programming for their comments on an earlier version of this paper.

## A LoC Breakdown of the Smart-Home Scenario Solutions

**Table 1** Overview of lines of code for the different scenarios according to the four identified concerns

| | openHAB | Hass | Elixir | Sparrow |
|---|---|---|---|---|
| State management | | | | |
| Automation 1 | 0 | 3 | 7 | 0 |
| Automation 2 | 4 | 2 | 10 | 0 |
| Automation 3 | 1 | 2 | 6 | 0 |
| Automation 4 | 3 | 2 | 5 | 0 |
| Automation 5 | 9 | 10 | 6 | 0 |
| Automation 6 | 1 | 1 | 10 | 0 |
| Automation 7 | 11 | 10 | 11 | 0 |
| Sequencing control | | | | |
| Automation 1 | 0 | 0 | 0 | 0 |
| Automation 2 | 0 | 0 | 0 | 0 |
| Automation 3 | 0 | 0 | 0 | 0 |
| Automation 4 | 0 | 0 | 0 | 0 |
| Automation 5 | 2 | 2 | 2 | 2 |
| Automation 6 | 0 | 0 | 0 | 0 |
| Automation 7 | 0 | 0 | 0 | 0 |
| Windowing management | | | | |
| Automation 1 | 0 | 0 | 0 | 0 |
| Automation 2 | 6 | 4 | 4 | 1 |
| Automation 3 | 6 | 1 | 2 | 1 |
| Automation 4 | 1 | 1 | 1 | 1 |
| Automation 5 | 4 | 6 | 2 | 2 |
| Automation 6 | 1 | 1 | 0 | 1 |
| Automation 7 | 2 | 2 | 3 | 1 |
| Pattern definition | | | | |
| Automation 1 | 3 | 2 | 5 | 2 |
| Automation 2 | 2 | 1 | 6 | 1 |
| Automation 3 | 4 | 1 | 4 | 1 |
| Automation 4 | 2 | 1 | 2 | 1 |
| Automation 5 | 6 | 3 | 4 | 6 |
| Automation 6 | 3 | 2 | 6 | 4 |
| Automation 7 | 4 | 3 | 4 | 1 |
| Total lines of code (LoC) | 75 | 60 | 100 | 25 |

# B Support of Synchronisation Operators by State-of-the-Art Proposals

**Table 2** Overview of the synchronization operators addressed in state-of-the-art proposals related to Sparrow

| | Filter Op. | | Selection Op. | Correlation Op. | | | Accum. Op. | | Transf. Op. |
|---|---|---|---|---|---|---|---|---|---|
| | Content-based | Time-based | Flexible | Conjunction | Disjunction | Sequencing | Count-based | Time-based | Aggregation |
| **Communication model extensions** | | | | | | | | | |
| ActorSpace[1] | x | - | - | - | - | - | - | - | - |
| TOTAM[45] | x | - | - | - | - | - | - | - | - |
| Directors[52] | x | - | - | - | - | - | - | - | - |
| Syndicate[20] | x | - | - | - | - | - | - | - | - |
| AErlang[12] | x | - | - | x[a] | - | - | - | - | - |
| interActors[21] | x | - | - | x[a] | - | - | - | - | - |
| RCF[22] | x | - | - | - | - | - | - | - | - |
| Coordinators[28] | x | - | - | - | - | - | - | - | x |
| **Monitor & Verification** | | | | | | | | | |
| Synchronizers[17] | x | - | - | x | x | - | - | - | - |
| RTSynchronizers[35] | x | x | - | x | x | x | - | - | - |
| Scoped-Synchronizers[13] | x | - | - | x | x | - | - | - | - |
| Moses[30] | x | - | - | - | - | - | - | - | - |
| ATC[29] | x | x | - | - | x | - | - | - | - |
| ARC[36, 46] | x | x | - | - | - | x | - | - | - |
| Ambient Contracts[44] | x | - | - | x[a] | - | x | - | - | - |
| MPSA[31] | x[b] | - | - | - | - | - | - | - | - |
| MPSA-Erlang[16] | x | - | - | - | - | - | - | - | - |
| lchanels[41] | x[c] | - | - | - | - | - | - | - | - |
| Effpi[42] | x[c] | - | - | - | - | x[d] | - | - | - |
| OTyPe[24] | x[b] | - | - | - | - | - | - | - | - |
| **Local synchronisation** | | | | | | | | | |
| Activators[18] | x | - | - | x[a] | x | - | - | - | - |
| Salsa[53] | x | - | - | x[a] | - | - | - | - | - |
| Reactive Isolates[34] | x | - | - | x[a] | - | - | - | - | x |
| AmbientTalk[50] | x | - | - | x[a] | - | - | - | - | - |
| Scala/Akka[49] | x | - | - | x[a] | - | - | - | - | - |
| Scala Joins[23] | x | - | - | x | - | - | - | - | - |
| JErlang[33] | x | - | - | x | - | - | - | - | - |
| Sparrow | x | x | x | x | x | x | x | x | x |

[a] Conjunction is only enforced to invocation of messages
[b] Additional type-based constraints are applied to message's attributes (e.g., MsgA(int, string))
[c] Type constraints to outgoing messages (object-level) are checked during the compilation phase
[d] Sequencing is only enforced to outgoing messages

## About the authors

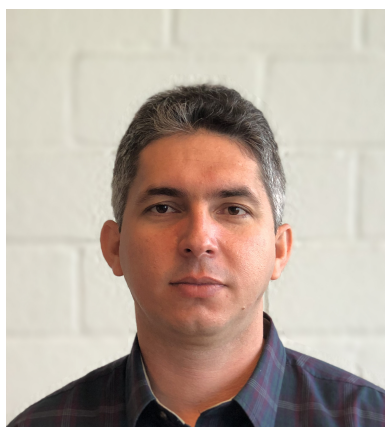

**Humberto Rodriguez Avila** is a PhD student at the Software Languages Lab, Vrije Universiteit Brussel in Belgium. His main research area is coordination of heterogeneous actors, and more concretely the design and implementation of programming techniques to express complex actor's interaction patterns. Contact him at rhumbert@vub.be.

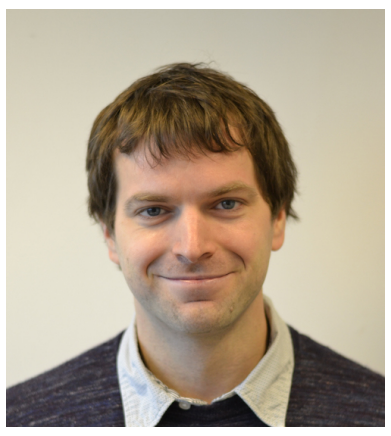

**Joeri De Koster** is an assistant professor in programming languages and runtimes. His current research is mainly focused on the design, formalisation and implementation of parallel and distributed programming languages. Contact him at jdekoste@vub.be.

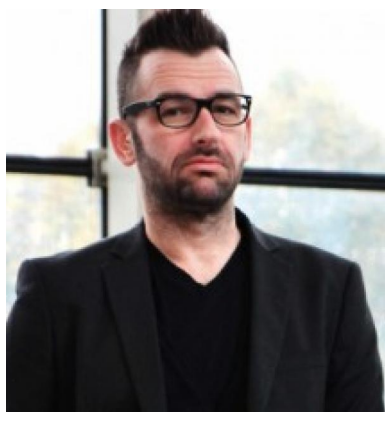

**Wolfgang De Meuter** is a professor in programming languages and programming tools. His current research is mainly situated in the field of distributed programming, concurrent programming, reactive programming and big data processing. His research methodology varies from more theoretical approaches (e.g., type systems) to building practical frameworks and tools (e.g., crowd-sourcing systems). Contact him at wdmeuter@vub.be.